\documentstyle[12pt]{article}
\newcommand{\sect}[1]{\section{#1}\setcounter{equation}{0}}

\begin{document}
\bigskip
\hspace*{\fill}
\vbox{\baselineskip12pt \hbox{UCSBTH-97-20}\hbox{hep-th/9709050}}
\bigskip\bigskip\bigskip

\centerline{\Large \bf Properties of Naked Black Holes}
\bigskip\bigskip

\centerline{\large Gary T. Horowitz and Simon F. Ross\footnote{\tt
gary@cosmic.physics.ucsb.edu, sross@cosmic.physics.ucsb.edu}}
\medskip
\centerline{Physics Department}
\centerline{University of California}
\centerline{Santa Barbara, CA 93106}
\bigskip\bigskip

\begin{abstract}
We investigate the properties of a class of near-extreme static black
hole solutions called naked black holes. These black holes, which
occur in string theory, have small curvature invariants but large
tidal forces outside their event horizons.  We show that these large
tidal forces are due to a concentration of dilaton stress-energy near
the horizon. We study infalling test strings and find that they are
highly excited by the large tidal forces, but remain small. Perhaps
most importantly, it turns out that a small amount of infalling matter
will cause the curvature invariants to become large outside the
horizon.  Nevertheless, an exact calculation shows that both the
matter trajectories and the classical black hole background are not
significantly altered.
\end{abstract}
 
 \newpage

\sect{Introduction}

We recently showed \cite{hor:naked} that there are black holes in
which the area of the event horizon is large, and all curvature
invariants are small near the horizon. Nevertheless, any object which
falls in experiences enormous tidal forces outside the horizon. The
tidal forces are given by the components of the Riemann tensor in a
frame associated with the ingoing geodesics. Thus, for these black
holes, the geodesic components are much larger than the invariants
constructed from them. This implies that the curvature is nearly
null. Although this is not the case for the familiar
Reissner-Nordstr\"om black hole, we found that it does occur in a wide
variety of theories involving scalar fields, including the low energy
limit of string theory. The examples were all charged black holes
either at or near extremality. These black holes were dubbed naked
black holes, as they violate the spirit of cosmic censorship, in that
large curvatures are visible outside of the horizon.

In this paper, we investigate the properties of these black holes in
more detail. We begin by trying to gain more insight into the nature
of the large curvatures seen by geodesic observers. We will see that
the size of the components of the Riemann tensor in the geodesic frame
is determined by the proper time remaining along the geodesic before
the singularity is reached. That is, the relevant length scale for the
geodesic curvature is this proper time, just as the relevant length
scale for the components in the more familiar static frame is set by
the area of the event horizon.  For the geodesics we consider, this
proper time will only be small for near-extreme black holes. This
makes it easier to understand why these black holes must always be
near-extreme.

We also show that the large curvature has a matter source. That is,
the large contributions to the Riemann tensor come from
the Ricci tensor, and hence from the stress-energy of the matter
fields. In our examples, we find that it is the dilaton fields'
stress-energy which provides the source. In fact, since the geometry
is spherically symmetric, only the dilaton field can have the nearly
null stress-energy needed to produce the type of curvature we
find. Hence, we would only expect to find black holes of this type in
theories with scalar fields. 

We then go on to consider the effects on infalling matter. If ordinary
matter falls into one of these black holes, it is crushed in the
transverse directions.  We study the effects on an infalling test string.
Considering the portion of the region of large curvatures that an
infalling string passes through, we find that it can be approximated
by a dilaton plane wave (under the assumption that the string remains
small compared to the horizon). This greatly simplifies the
calculation, as the propagation of strings through such plane waves
has been studied previously \cite{hor:wave,vesa:sing}.  We find that
the string becomes very excited before it crosses the horizon. Thus
its average mass becomes large. However, the average size remains
small, as the modes are all oscillating very quickly near the
horizon. In this context, we also consider the effect on outgoing
strings which form part of the Hawking radiation.

Perhaps the most important effect of infalling matter is seen when we
go beyond the test particle (string) approximation and include its
stress-energy. It then turns out that a small amount of matter can
produce Planck-scale invariants outside the horizon.  This can be
viewed as the result of a high energy collision between the infalling
matter and the background dilaton wave.  Surprisingly, by carrying out
an exact calculation of the collapse of a charged dust shell, we find
that these large invariants do not significantly change the classical
trajectory of the infalling matter. Furthermore, after the shell
passes, the spacetime is simply a charged black hole with slightly
larger charge and mass. So the large curvature invariants have little
effect on the classical evolution.  In fact we will see that there is
a sense in which certain spacetimes with large curvature invariants
are ``close" to spacetimes with small invariants.  The main
consequence of the Planck scale curvature invariants is that quantum
effects will become important outside the horizon when matter falls
in.

In the next section, we review the results of \cite{hor:naked},
discussing two examples which will be used later in this paper.  In
section \ref{source}, we discuss the relation between the tidal forces
and the proper time remaining before the observers reach the
singularity. We also show that the large curvature is associated with
a large dilaton stress tensor. In section \ref{effects}, we consider
infalling matter, and show that the geometry as seen by small
infalling observers is approximately that of a dilaton plane wave. We
go on to show that infalling test strings will be highly excited.  In
section \ref{backr}, we consider the stress energy due to infalling
matter, and show that it can lead to large invariants. We then
demonstrate that these do not imply large changes in either the
classical solution or the motion of the matter.  A brief discussion is
contained in section \ref{concl}.

\sect{Review of naked black holes}
\label{rev}

In \cite{hor:naked}, we showed that extremal or near-extremal limits
of several familiar black hole solutions arising in general relativity
with scalar matter fields or in string theory have large tidal forces
outside large event horizons.  We will focus here on four dimensional
black holes with metrics of the form
\begin{equation} \label{gmet}
ds^2 = -{F(r) \over G(r)} dt^2  + {dr^2 \over
F(r)} + R^2(r) d\Omega.
\end{equation}
This metric will have a horizon at $r=r_+$ if $F(r_+)=0$. The usual
static frame is
\begin{eqnarray} \label{sframe}
(e_0)_\mu &= &-F^{1/2}(r)G^{-1/2}(r)\ \partial_\mu t,
\qquad (e_1)_\mu = F^{-1/2}(r)\ \partial_\mu r, \\ 
(e_2)_\mu &=& R(r)\ 
 \partial_\mu\theta,  \quad\qquad\qquad\qquad
(e_3)_\mu = R(r)\sin\theta\ \partial_\mu \phi. \nonumber
\end{eqnarray} 
We want to compare the curvature components in this frame with the
physical tidal forces felt by infalling observers. Thus, we consider a
frame which is parallelly propagated along radially infalling
geodesics. The geodesics have tangent vector $u = (\dot{t},
\dot{r},0,0)$, where a dot denotes $d/d\tau$. There is a conserved
energy per unit mass $E = \dot{t} F(r)/G(r)$, and
\begin{equation} \label{rdot}
\dot{r}^2 =  E^2 G(r) - F(r).
\end{equation} 
We will always assume that $E$ is of order one,
i.e., we will consider geodesics that start at infinity with small
velocity.  The parallelly propagated orthonormal frame, in which
$(e_{0'})_\mu = u_\mu$, is then related to the static frame by a
radial boost,
\begin{eqnarray} \label{inframea}
(e_{0'})_\mu &=& u_\mu = -E \partial_\mu t + {\dot{r}
\over F(r)} \partial_\mu r \\  &=& \cosh \alpha
(e_0)_\mu + \sinh \alpha (e_1)_\mu, \nonumber
\end{eqnarray}
and 
\begin{equation} \label{inframeb}
(e_{1'})_\mu = \sinh \alpha (e_0)_\mu + \cosh \alpha (e_1)_\mu, 
\end{equation} 
where $\cosh \alpha = E [G(r)/F(r)]^{1/2}$. Note that since the
horizon lies at $F(r) = 0$, the boost parameter $\alpha$ diverges as
we approach the horizon. The curvature components $R_{0'k0'k}$,
$k=2,3$, can be calculated by using the geodesic deviation equation,
with the result \cite{hor:naked}
\begin{equation}\label{gdevb}
R_{0'k0'k} = -{\ddot{R}\over R}.
\end{equation}
Although the boost parameter diverges at the horizon, these curvature
components will generally be finite. However, in certain cases they
can be much larger than the curvature components in the static frame.

The dilaton black holes discussed in \cite{gib:bh1,gib:bh2,gar:chbh}
are the simplest examples. These are solutions of a theory with a
Maxwell field $F_{\mu\nu}$ and a scalar field $\phi$ with the coupling
between the Maxwell field and the scalar field governed by an
arbitrary constant $a$. The action is
\begin{equation}\label{dbhac}
S = \int d^4 x \sqrt{-g} \left[R - 2 (\nabla
\phi)^2 - e^{-2a\phi} F_{\mu\nu}F^{\mu\nu}\right],
\end{equation}
and the metric for a dilaton black hole  is given by
\begin{equation}\label{gensol}
ds^2 = -F(r) dt^2 + {dr^2 \over F(r)} + R^2(r)
d\Omega,
\end{equation}
with
\begin{equation}\label{uuf}
F(r) = {(r-r_+)(r-r_-) \over R^2}
\end{equation}
and
\begin{equation}\label{R}
R(r) = r \left( 1- {r_- \over r} \right)^{a^2/(1+a^2)}.
\end{equation}
The dilaton is given by
%
%
\begin{equation} \label{dbhdil}
e^{-2\phi} = \left( 1 - \frac{r_-}{r} \right)^{2a/(1+a^2)}
\end{equation}
if the solution carries a magnetic charge. There is a horizon at
$r=r_+$ and a singularity at $r=r_-$ for $a \neq 0$. For $a=0$, this
metric reduces to the Reissner-Nordstr\"om metric; $r=r_-$ is an inner
horizon, and there is a singularity at $r=0$.  The extremal limit in
both cases is $r_+ = r_-$.  The ADM mass and charge are
\begin{equation}\label{dbhmass}
M={r_+\over2} + \left({1-a^2\over 1+a^2}\right) {r_-\over 2},
\end{equation}
\begin{equation}
Q=\left({r_+ r_-\over 1+a^2}\right)^{1/2}. 
\end{equation}
The Hawking temperature for these black holes is
\begin{equation}
T = \frac{1}{4\pi} \frac{(r_+ -r_-)^{1-a^2 \over 1+a^2}}{r_+^{2 \over
1+a^2}}.
\end{equation}

The horizon area will be large and the static curvature will be small
(in Planck units) if
\begin{equation}\label{dbhr}
R(r_+) = r_+\epsilon^{a^2/(1+a^2)} \gg 1,
\end{equation}
where $\epsilon\equiv (1-r_-/r_+)$.  Note that the exponent of
$\epsilon$ is always less than one. The curvature in a freely falling
frame is given by
\begin{equation}\label{bhtide}
R_{0'20'2} = -{\ddot{R}\over R}= -{1 \over R}\left[ R''(E^2 - F) - {F'
R' \over 2}\right]. 
\end{equation}
Near the horizon, $F(r)$ is small, so this will be larger than the
Planck scale if
\begin{equation}\label{dbhtide}
 \left|{R'' \over R} \right|
= {a^2 \over (1+a^2)^2} {(1-\epsilon)^2 \over r_+^2
\epsilon^2}>1.
\end{equation}
This will be satisfied, for $a \neq 0$, if $r_+ \epsilon \ll 1$. Thus
we see that there is a range of parameters for which the curvature in
the static frame is small, but infalling observers experience large
tidal forces near the horizon, namely $\epsilon \ll 1$ and
$\epsilon^{-a^2/(1+a^2)} \ll r_+ \ll \epsilon^{-1}$. Since $\epsilon$
is small, these black holes are all close to extremality, and since
$r_+$ is large, they have a large mass. For fixed mass, the area of
the event horizon goes to zero in the extremal limit. The spacetime
develops a null singularity if $0<a \le 1$ and a timelike singularity
if $a>1$.  We are considering a different limit, in which the mass is
increased as one approaches extremality, so the horizon area remains
large. Note that 
\begin{equation} \label{tempext}
(r_+ - r_-) =  4\pi T R(r_+)^2.
\end{equation}
Thus, if we take $r_+ \to r_-$ while keeping the horizon area fixed, the
Hawking temperature will go to zero.

In the context of string theory, the black hole solution
with electric Neveu-Schwarz charges associated with internal momentum
and string winding number has similar behavior. 
The string metric is \cite{sen:bh}
\begin{equation}\label{nsmet}
ds^2 = - \Delta^{-1} \left(1-{r_0\over r}\right) dt^2 + 
\left(1-{r_0\over r}\right)^{-1}
dr^2 + r^2 d\Omega,
\end{equation}
where 
\begin{equation}\label{dd}
\Delta = \left( 1+ {r_0 \sinh^2 \gamma_1 \over r} \right) \left(
1 + {r_0 \sinh^2 \gamma_p \over r} \right),
\end{equation}
and the dilaton is given by
\begin{equation}\label{nsdil}
e^{2 \phi} = \Delta^{-1/2}.
\end{equation}
The ADM mass of these black holes is
\begin{equation}\label{nsmass}
M = {r_0 RV \over g^2}(2+\cosh 2\gamma_1 + \cosh
2\gamma_p),
\end{equation}
and the integer normalized charges are 
\begin{equation}\label{nscharge}
n = {R^2 V \over g^2} r_0 \sinh 2 \gamma_p, \qquad m =
{V \over g^2} r_0 \sinh 2 \gamma_1,
\end{equation}
where $R$ is the radius of a compact internal direction, and 
$(2\pi)^5 V$ is the volume of an internal five-torus. (We are using
the same conventions  as \cite{mald:D-ent} and have set $\alpha' =1$.)

The curvature in the static frame is of order $1/r_0^2$ at the horizon
$r=r_0$, so we must take $r_0 \gg 1$ to keep it small. The
curvature in the infalling frame at $r=r_0$ is
\begin{equation}\label{nsgeod}
R_{0'20'2} = -{\ddot{R} \over R} = -{R' \over 2 R} \left(
G' E^2 - F' \right) = -{E^2 \Delta' \over 2 r_0} + {1 \over 2 r_0^2}.
\end{equation}
Since 
\begin{equation}\label{ddd}
\Delta'(r_0) = -{1 \over r_0}(\sinh^2 \gamma_1 \cosh^2 \gamma_p +
 \cosh^2 \gamma_1 \sinh^2 \gamma_p ),
\end{equation}
we can make the tidal forces arbitrarily large by increasing
$\gamma_1$ or $\gamma_p$.  Physically, this just corresponds to
increasing the mass and charges.

The extremal limit for this class of black holes is $r_0 \to 0$,
$\gamma_1, \gamma_p \to\infty$ with $n, m$ fixed. It may appear that
the large tidal forces are present far from the extremal limit, since
we have taken $r_0 \gg 1$. However, for fixed charges, the mass above
extremality is
\begin{equation}\label{dele}
\Delta M = M - M_{ext} \approx {2 r_0 RV \over g^2}.
\end{equation}
When $r_0 \gg 1$, $\Delta M$ is large, but $\Delta M/M$ is still small,
since $\gamma_1$ or $\gamma_p$ is large.

The Einstein metric is obtained by multiplying (\ref{nsmet}) by
$e^{-2\phi}$.  The area of the horizon is thus increased by the factor
$\cosh\gamma_1 \cosh\gamma_p$ and the static frame curvature near the
horizon is decreased by the same factor. If one takes $r_0$ small and
$\gamma_1,\gamma_p$ sufficiently large, then both the size of the
black hole and the tidal forces in the Einstein metric will be
large. This example is closely related to the dilaton black hole
case. If $\gamma_1 = \gamma_p$, the Einstein metric is the same as the
dilaton black hole with $a=1$. If $\gamma_1=0$ or $\gamma_p=0$, the
metric is the same as the dilaton black hole with $a=\sqrt 3$.

\sect{Source of the large curvature}
\label{source}

In this section, we will discuss the origin of the large geodesic
curvatures found in naked black holes. One of the first questions one
might ask about these curvatures is what sets their size. That is, in
the dilaton black hole spacetime, there is a natural large distance
scale, namely the radius $R(r_+)$ of the black hole horizon.  This
scale determines the size of the curvature in the static frame. Can we
identify a short distance scale that determines the size of the
curvature components in the geodesic frame?  For the simpler examples,
the answer is yes. The relevant scale is the proper time remaining
along the geodesics before the observers reach the
singularity\footnote{This has also been noticed by B. Hiscock and
S. Larsen (private communication). For the Reissner-Nordstr\"om
metric, timelike geodesics never hit the singularity, but that is not
true for the black holes considered here.}.

Consider the dilaton black hole metrics. Near the horizon, $F(r)$ is
negligibly small, so we can see from (\ref{rdot}) that $\dot{r}^2
\approx E^2$, and hence
\begin{equation}
\tau \approx (r-r_-)/E,
\end{equation}
where we have chosen the sign to be positive, so that $\tau$ corresponds
to the `time remaining', and decreases as we follow the geodesic into
the black hole. Note that although we have chosen the constant of
integration so that the left-hand side vanishes at the singularity,
this expression is not valid for $r$ near $r_-$ if $a^2 >1$, as in that
case $F(r)$ will eventually diverge as we approach the
singularity. Thus, in that case the actual proper time remaining will
be somewhat shorter. For $a^2 \leq 1$, the expression is valid up to
the singularity. 

Now consider the function $R(r)$ in (\ref{R}). At the horizon,
\begin{equation}
R(r_+) = r_+^{1 \over 1+a^2}(r_+ - r_-)^{a^2 \over 1+a^2}.
\end{equation} 
Since we are considering the near-extreme case where $r_+ - r_-
\ll 1$, if we move slightly away from the horizon, the second term
will vary much more quickly than the first term. Thus
\begin{equation} \label{Rtau}
R(r) \approx  r_+^{1 \over 1+a^2}(r - r_-)^{a^2 \over 1+a^2} \propto
\tau^{a^2 \over 1+a^2} 
\end{equation}
in a neighborhood of the horizon, and hence the curvature is
\begin{equation} \label{curvtau}
R_{0'20'2} = -{\ddot{R}\over R} \approx \frac{a^2}{(1+a^2)^2}
\frac{1}{\tau^2}. 
\end{equation}
Thus, near the horizon, the size of the curvature in the geodesic frame
is set by the proper time remaining to the singularity.
The fact that the factor in front of $1/\tau^2$ in
this expression is independent of the black hole's mass and charge
indicates that this is a useful way to characterise the curvature.

For the Neveu-Schwarz black holes, the situation is somewhat more
complicated, as we have one more parameter. If both of the charges are
large, we find that $R_{0'k0'k} = \frac{1}{4 \tau^2}$, and if one of
the charges is zero, $R_{0'k0'k} = \frac{2}{9 \tau^2}$. If both
charges are non-zero but one is much smaller than the other, we cannot
give a simple form for the curvature that is applicable throughout the
region of large curvature. However, it is still true that the proper
time remaining along the geodesics is short, and hence this may still
be the relevant distance scale.

It also worth pointing out that the simple expression obtained above
applies only to the large curvatures experienced near the horizon by
geodesic observers falling from rest at infinity. In the dilaton black
hole, even if we are far from the black hole, we can always find
geodesic observers (with sufficiently large $E$) who see large
curvatures. However, the analogue of (\ref{curvtau}) for such an
observer far from the black hole is $R_{0'k0'k} \propto
\frac{r_-^2}{r^2 \tau^2}$, where $r$ is the radial distance to the
black hole. So while we can still make the curvature large at any $r$
by taking $\tau$ small enough, $\tau$ doesn't set its scale.

The fact that the proper time remaining sets the scale of the
curvature makes it easier to understand why such large curvatures have
only been found for near-extreme black holes. It is only for such
black holes that the proper time remaining at the horizon is short for
observers who have fallen in from infinity. This makes it seem
unlikely that more general black hole solutions will exhibit similar
behaviour, even once we include quantum corrections.

Although it is useful to identify the proper time remaining as the
relevant length scale, we still need to ask what the source of the
large curvatures is. One can decompose the Riemann tensor in terms of
the Weyl tensor and the Ricci tensor. For static spherically
symmetric black hole solutions, we now show that the Weyl tensor is
always invariant under radial boosts. Since the difference between
the static frame and the infalling frame is just such a boost, this
shows that the difference in curvature must come from the Ricci tensor,
and hence the matter fields.

In four spacetime dimensions, one can show that the Weyl tensor is
boost invariant by using the fact that every static spherically
symmetric spacetime is algebraically special of type D
\cite{wald:gr}. The repeated principal null directions are just the
radial ingoing and outgoing null vectors. In two component spinor
notation, this means that the Weyl spinor can be expressed as
\begin{equation}
\Psi_{ABCD} = \Psi o_{(A} o_B \iota_C\iota_{D)}.
\end{equation}
Since $o_A \to \lambda o_A, \ \iota_A \to \lambda^{-1} \iota_A$ under
a boost, the Weyl spinor is clearly invariant, and hence so is the
Weyl tensor.

In $d$ spacetime dimensions, one can establish the same result as
follows.  The Weyl tensor can be written as
\begin{equation} \label{weyl}
C_{\mu\nu\rho\sigma} = R_{\mu\nu\rho\sigma} - \frac{2}{d-2}
(g_{\mu[\rho} R_{\sigma]\nu} - g_{\nu[\rho} R_{\sigma]\mu}) +
\frac{2}{(d-1)(d-2)} R g_{\mu[\rho} g_{\sigma]\nu}. 
\end{equation}
Under a radial boost,
\begin{equation} \label{bcurva}
R_{0'1'0'1'} = R_{0101},\qquad  R_{0'k1'k} = \cosh
\alpha \sinh \alpha (R_{0k0k} +R_{1k1k}),
\end{equation}
\begin{equation} \label{bcurvb}
 R_{0'k0'k} = R_{0k0k} + \sinh^2 \alpha (R_{0k0k} + R_{1k1k}),
\end{equation}
\begin{equation} \label{bcurvc}
R_{1'k1'k} = R_{1k1k} + \sinh^2 \alpha (R_{0k0k} + R_{1k1k}),
\end{equation}
and $R_{klkl}$ is unchanged, where $\alpha$ is the boost parameter. It
follows that 
\begin{equation} \label{briccia}
R_{0'0'} = R_{0'1'0'1'} + \sum_k R_{0'k0'k} = R_{00} +  \sinh^2
\alpha\sum_k  (R_{0k0k} + R_{1k1k}) 
\end{equation}
and
\begin{equation} \label{briccib}
R_{1'1'} = -R_{0'1'0'1'} + \sum_k R_{1'k1'k} = R_{11} +  \sinh^2
\alpha\sum_k  (R_{0k0k} + R_{1k1k}), 
\end{equation}
and that $R_{kk}$ and $R_{11} - R_{00}$ are boost invariant. It is
then trivial to check that $C_{0101}$ and $C_{klkl}$ are boost
invariant. We have 
\begin{eqnarray} \label{bweyla}
C_{0'k0'k} &=& R_{0'k0'k} - \frac{1}{d-2} (-R_{kk} + R_{0'0'}) -
\frac{2}{(d-1)(d-2)} R \\ &=& C_{0k0k} + \sinh^2 \alpha \left[
(R_{0k0k} + R_{1k1k}) - \frac{1}{d-2} \sum_l(R_{0l0l} +R_{1l1l})
\right],   \nonumber 
\end{eqnarray}
and similar expressions for $C_{1'k1'k}$ and $C_{0'k1'k}$. Since $l$
in the summation runs over $d-2$ values, the Weyl tensor will be
boost-invariant if all the $R_{0l0l} + R_{1l1l}$ are the same.

For a spherically symmetric black hole solution, $l$ just runs over
the angular variables, and the spherical symmetry thus implies that
all the terms are the same. Thus, for both the dilaton and
Neveu-Schwarz examples, the Weyl tensor is
boost-invariant\footnote{This is not the case for the black $p$-brane
solutions.}. The large value of the Riemann tensor in the geodesic
frame is entirely due to the Ricci tensor.

The Ricci tensor is determined by the stress-energy tensor of the
matter fields. Thus, the large curvatures are not mysterious at all;
they are simply the consequence of a large matter source near the
horizon.  For a radial electric field, $ \mbox{ \boldmath $F$} \propto
(e_0) \wedge (e_1)$, while for a radial magnetic field, {\boldmath $F
\propto \epsilon$}, where {\boldmath $\epsilon$} is the volume element
on $S^2$, and both of these are unchanged by radial boosts. Any static
spherically symmetric electromagnetic field $F_{\mu\nu}$, and hence
the stress-energy tensor associated with it, is therefore always
boost-invariant. Thus, the large stress-energy must be associated with
the dilaton field. This shows that black holes of this type will only
occur in theories with a scalar field.

The curvature is becoming large and nearly null near the horizon. We
therefore see that the source of the large geodesic curvatures is a
large, nearly null dilaton stress tensor near the horizon. This is
closely analogous to the dilaton plane waves studied in
\cite{hor:wave}. The symmetries are also similar, since the timelike
Killing field is becoming null near the horizon, and the spherical
symmetry implies invariance in the transverse directions.  In fact, as
we will see in the next section, the region of the geometry explored
by small infalling observers is well approximated by such a plane
wave.

To conclude this section, consider the effects of large tidal forces
on some classical object falling into the black hole. Once the object
is in the region where the tidal forces are large, they will dominate
over any internal stresses in the body (since such stresses can
reasonably be assumed to be very small compared to the Planck scale).
Thus, the evolution of the object is well-approximated by regarding it
as a cloud of dust, and is characterised by how much this dust cloud
is distorted in passing through the region of large tidal forces, up
to the horizon. Since the solutions are spherically symmetric, the
distortion will be just a uniform shrinking in the transverse
directions. In other words, the object is crushed.  The amount of
shrinking is simply given by the change in $R(r)$ as we pass through
the region of large tidal forces. In our examples, this is always
finite, but not necessarily small.

\sect{Effects on test strings}
\label{effects}

In this section, we will consider the effects of the large tidal
forces on infalling test strings. As we have argued above, the
spacetime in a neighborhood of the horizon is closely analogous to one
of the plane wave spacetimes studied in \cite{hor:wave}. We now show
explicitly that if we consider the part of the spacetime traversed by
an observer of small spatial extent whose center of mass follows a
geodesic in from infinity, it can in fact be approximated by one of
these plane waves in the region of large curvatures outside the
horizon. Since we are interested in the effect on infalling strings,
we use the string metric describing a Neveu-Schwarz charged black
hole. We will assume that both the winding and momentum charges are
large, as this is the simplest case.

First we need to introduce Kruskal-like coordinates for the black hole
solution. We define a tortoise coordinate $r_*$ such that 
\begin{equation} \label{tort}
dr_* = \Delta^{1/2}\left( 1 - \frac{r_0}{r} \right)^{-1} dr .
\end{equation}
We then define the Kruskal-like coordinates $U = -e^{-\kappa
  (t-r_*)}$, $V = e^{\kappa (t+r_*)}$, where $\kappa$ is the surface
gravity of the black hole. The metric in terms of
these new coordinates is
\begin{equation} \label{krusmet}
ds^2 = {\left( 1 - \frac{r_0}{r} \right) \over \Delta \kappa^2 UV} dU
dV + r^2 d\Omega,
\end{equation}
where $r$ is given in terms of $U$ and $V$ by $UV = -e^{2\kappa r_*}$. 

We are considering the case where both charges are large, $\gamma_1,
\gamma_p \gg 1$. Therefore, so long as $r^2 \ll r_0^2 \sinh^2 \gamma_1
\sinh^2 \gamma_p$ (that is, throughout the region of large curvatures),
\begin{equation} \label{appD}
\Delta \approx {r_0^2 \sinh^2 \gamma_1 \sinh^2 \gamma_p \over r^2},
\end{equation}
and hence 
\begin{equation} \label{apprs}
r_* \approx r_0 \sinh \gamma_1 \sinh \gamma_p \ln \left( {r-r_0 \over
r_0} \right),
\end{equation}
where we have chosen the constant of integration so that $r_*= 0 $ at
$r= 2r_0$. Further, $\kappa \approx 1/(2r_0 \sinh \gamma_1 \sinh
\gamma_p )$, so 
\begin{equation} \label{appmet}
ds^2 \approx -4 r_0 r dU dV + r^2 d\Omega,
\end{equation}
and
\begin{equation} \label{appUV}
UV \approx - \left( {r-r_0 \over r_0} \right).
\end{equation}

For radial geodesics falling into the black hole from infinity,
(\ref{rdot}) and $\dot{t} = -E \Delta/(1-r_0/r)$ imply 
\begin{equation} \label{rsdot}
\dot{r}_* = \Delta^{1/2}\frac{\dot{r}}{\left(1- {r_0 \over r} \right)}
= \frac{\Delta^{1/2}}{\left(1- {r_0 \over r} \right)} \left[E^2 \Delta
- \left(1- {r_0 \over r} \right)\right]^{1/2} \approx -\dot{t} \left[ 1
- \left(1- {r_0 \over r} \right) {1 \over 2 E^2 \Delta  } \right],
\end{equation}
where the approximation is again valid for the region of large
curvatures, and an overdot denotes the derivative with respect to the
proper time remaining before the singularity is reached. It follows that
\begin{equation} \label{Vdot}
\dot{V} = V \kappa (\dot{t} + \dot{r}_*) \approx V \kappa \dot{t} \left(1-
{r_0 \over r} \right) {1 \over 2E^2 \Delta } = -{V \kappa \over 2 E},
\end{equation}
so $|\dot{V}| \ll V$, as $\kappa \ll 1$. Thus, infalling observers who
follow the geodesics have $V$ nearly constant in the region of large
curvatures. Note also that
\begin{equation} \label{Udot}
\dot{U} = -U \kappa (\dot{t} - \dot{r}_*) \approx -U 2 \kappa \dot{t}.
\end{equation}
We now set
\begin{equation} \label{Vo}
V = V_0 \left( 1- {v_1 \over 2r_0^2} \right),
\end{equation}
and ignore the $v_1$ dependence in the metric. We can then write 
\begin{equation} \label{ru}
r \approx r_0 (1 - U V_0),
\end{equation}
and
\begin{equation} \label{metu}
ds^2 \approx 2 (1-U V_0) V_0 dU dv_1 + (1-U V_0)^2 r_0^2 d\Omega. 
\end{equation}

It is convenient to define $u = (1-UV_0)^2$, so the horizon
corresponds to $u=1$. Since the event horizon is large, and the
infalling observer is assumed to be small, we can also approximate the
two-sphere metric $r_0^2 d\Omega$ by a flat metric. We will use
$X_1$ and $X_2$ to denote these flat directions.
We can now rewrite the metric
as
\begin{equation} \label{dsgary}
ds^2 \approx -du dv_1 + u dX_i dX^i. 
\end{equation}
where $i=1,2$. This is a plane wave metric. To bring it into the form
used in \cite{hor:wave} we use a change of coordinates discussed in
\cite{gib:wave},
\begin{equation} \label{vc}
v = v_1 + \frac{1}{2} X_i X^i, 
\end{equation}
\begin{equation} \label{xc}
x_i = \sqrt{u} X_i.
\end{equation}
Then the metric is
\begin{equation} \label{pwmet}
ds^2 \approx -du dv + dx_i dx^i+ W(u) x_i x^i du^2 ,
\end{equation}
where
\begin{equation} \label{W}
W(u) = -{1 \over 4 u^2}.
\end{equation}
Since the horizon corresponds to $u=1$, it might seem from this form
of the metric that the tidal forces are not large there. However, an
observer who falls into the black hole with energy $E$ at infinity
has
\begin{equation} \label{P}
P \equiv \dot{u} = -2(1-UV_0) V_0 \dot{U} \approx 4(1-UV_0) UV_0 \kappa
\dot{t} \approx 4 {r^2 \over r_0^2} \kappa \Delta E \approx 2 {\sinh
\gamma_1 \sinh \gamma_p \over r_0} E \gg 1.  
\end{equation}
Thus, these observers experience large tidal forces, as they are
passing through the wave very quickly. This plane wave approximation
is valid throughout the region of large curvatures outside the
horizon, which is presumably the region in which interesting effects
on strings will be produced.  One can think of the singularity as
being at $u=0$, although the plane wave approximation is no longer
valid there, as we cannot treat the two-sphere as flat once we are
sufficiently close to the singularity.

The propagation of first quantized strings through plane waves was
discussed in detail in \cite{hor:wave}, and we will use the same
approach and conventions. However, there is one modification. In
\cite{hor:wave}, the main interest was in considering the effects of
spacetime singularities on strings, so the spacetime considered
consisted of a non-trivial wave with flat spacetime regions before and
after it. In our case, we are interested in the behaviour of the
string in the region near the horizon, so it is not appropriate to
match onto a flat region beyond the wave. However, to interpret the
results, we need a static region `after' the wave where we can define
positive and negative frequencies. We will therefore consider a
spacetime of the form (\ref{pwmet}), with $W$ given by (\ref{W}) up to
the horizon $u = 1$, and then match onto $W = - 1/4 $ in the region $u
< 1$. Even though the resulting metric component $W$ is only
continuous and not differentiable, there is no induced stress-energy
along the matching surface. This is because a metric of the form
(\ref{pwmet}) has $R_{\mu\nu} = -W \partial_\mu u \partial_\nu u$.

Since the spacetime is a plane wave, we can choose the light-cone
gauge for the string. That is, we choose coordinates $(\sigma,\tau)$
on the worldsheet so that $u = P\tau$. We assume $P\gg 1$ since an
unexcited string which falls in from rest at infinity will follow an
approximate geodesic until it reaches the region of large tidal
forces. The dynamical fields are $x^i (\sigma,\tau)$, $i=1,2$. If we
decompose the $x^i$ into modes,
\begin{equation} \label{xmode}
x^i(\sigma,\tau) = \sum_n x^i_n(\tau) e^{in\sigma},
\end{equation}
then the worldsheet field equation for $x^i$ becomes
\begin{equation} \label{xeq}
\ddot{x}^i_n + n^2 x^i_n -WP^2 x^i_n = 0,
\end{equation}
where a dot denotes differentiation with respect to $\tau$. Since we
have modified the metric at the horizon $u=1$ ($\tau = 1/P$)\footnote{Recall
that $\tau$ denotes the time remaining to the singularity. So $\tau$
decreases as the string falls in.},
\begin{equation}
WP^2 = -{1\over 4\tau^2} \quad {\rm for} \quad \tau > {1\over P},
\nonumber 
\end{equation}
\begin{equation}
WP^2 = -{P^2\over 4} \quad {\rm for} \quad \tau <  {1\over P}.
\end{equation}
The component $v(\sigma,\tau)$ is determined by
\begin{equation} \label{v1}
P\dot{v} = (\dot{x}_i)^2 + (x_i')^2 +WP^2x^2,
\end{equation}
\begin{equation} \label{v2}
P v' = 2\dot{x}_i x_i',
\end{equation}
where a prime denotes differentiation with respect to
$\sigma$. 

We can write the solutions of the mode equation (\ref{xeq}) in terms
of a complete set of solutions which are pure positive and negative
frequency asymptotically. That is, 
\begin{equation} \label{asexp}
x^i_n = i(a^i_n u_n - \tilde{a}^{i^\dagger}_n \tilde{u}_n),
\end{equation}
where $u_n$ and $\tilde{u}_n$ are solutions of (\ref{xeq}) and
\begin{equation} \label{un}
u_n \to \frac{1}{2\sqrt{ n}}e^{-in\tau} , \qquad \tilde{u}_n \to
\frac{1}{2\sqrt{ n}}e^{in\tau} 
\end{equation}
when $\tau \to \infty$.
 Similarly, we can
write the solution in terms of a set of solutions which are positive and
negative frequency in the region $\tau < 1/P$. That is,
\begin{equation} \label{postexp}
x^i_n = i(b^i_n v_n - \tilde{b}^{i^\dagger}_n
\tilde{v}_n), 
\end{equation}
where $v_n$ and $\tilde{v}_n$ are solutions of (\ref{xeq}) which are
\begin{equation} \label{vn}
v_n = \frac{1}{2\sqrt{ \tilde{n}}}e^{-i\tilde{n}\tau},\qquad \tilde{v}_n
= \frac{1}{2\sqrt{ \tilde{n}}} e^{i\tilde{n}\tau}
\end{equation}
for $\tau <1/P$, where
\begin{equation}\label{ntdle}
\tilde{n}^2 = n^2 +{P^2\over 4}. 
\end{equation}
Since both the $u$'s and the
$v$'s are complete sets of states, it must be possible to write them
in terms of each other. This implies a linear transformation between
the initial and final mode creation and annihilation operators, the
Bogoliubov transformation
\begin{equation} \label{bog}
b^i_n = A_n a^i_n - B^*_n \tilde{a}^{i^\dagger}_n, \qquad
\tilde{b}^i_n = A_n \tilde{a}^i_n - B^*_n a^{i^\dagger}_n. 
\end{equation}
The mass-squared and average size of a string initially in the ground
state are related to the Bogoliubov coefficients $B_n$. 

A string in the static region $\tau <1/P$ is similar to one in flat
spacetime except that the gravitational field has shifted the
frequency of the $n^{th}$ mode from $n$ to $\tilde n$
(\ref{ntdle}). In particular, the mass of the string is given by
\begin{equation}\label{smass}
M^2_s = 8\sum_{n=1}^\infty \tilde n  {b^i_n}^\dagger b^i_n  .
\end{equation}
This follows by integrating (\ref{v1}) over $\sigma$.
We have ignored the usual normal ordering constant, since that will be
small compared to the mass of the excited string.

We need to calculate the Bogoliubov coefficient $B_n$ between the
asymptotically flat region and the region $\tau < 1/P$.  Since the
$u_n$'s and $v_n$'s are normalized, we can calculate this coefficient
by evaluating the inner product between the initial positive-frequency
mode $u_n$ and the final negative-frequency mode
$\tilde{v}_n$. Fortunately, (\ref{xeq}) has a simple closed-form
solution in the region $\tau > 1/P$ \cite{vesa:sing}. It is
\begin{equation} \label{soln}
x_n = \sqrt{\tau} \left[ A J_0(n\tau) + B N_0(n\tau)\right],
\end{equation}
where $A$ and $B$ are arbitrary constants, and $J_0$ and $N_0$ are the
Bessel functions of the first and second kinds. For $u_n$, the
boundary condition (\ref{un}) and the asymptotic expansions of the
Bessel functions imply that
\begin{equation} \label{un2}
u_n = \sqrt{\frac{\pi \tau}{8}} e^{-i\pi/4} [J_0(n\tau) - i
N_0(n\tau)]. 
\end{equation}
We expect that only the modes with $n \ll P$ will be
significantly excited by the wave, as it is for these modes that the
coefficient of $x^i_n$ in (\ref{xeq}) changes significantly as we
pass through the wave.  We therefore calculate the Bogoliubov
coefficients only for $n \ll P$. In this case, we can use the
short-distance expansion of the Bessel functions near the matching
point at $\tau = 1/P$. That is, for $\tau \sim 1/P$,
\begin{equation} \label{unmat}
u_n \approx \sqrt{\frac{\pi \tau}{8}} e^{-i\pi/4} \{1 - 2i [{\bf C} + \ln
(n\tau/2) ]\},
\end{equation}
where {\bf C} is Euler's constant. For $\tau<1/P$, the solution
will take the form 
\begin{equation} \label{vnmat}
u_n= T \frac{1}{2\sqrt{ \tilde{n}}} e^{-i\tilde{n}\tau} + R
\frac{1}{2\sqrt{ \tilde{n}}}e^{i\tilde{n}\tau}.
\end{equation}
The constants $T$ and $R$, which are determined by matching the two
forms at $\tau = 1/P$, are essentially the Bogoliubov coefficients.
Performing this matching, we find that $B_n \propto \ln (n /P)$.

Thus if a string is initially in its ground state and falls into the
black hole, by the time it reaches the horizon, each mode with $n 
 \ll P$ will be excited to $\langle N_n\rangle \equiv \langle 0|
{b_n^i}^\dagger b_n^i |0\rangle \sim \ln^2(n /P)$. The excitation in
modes $n \gg P$ will be highly suppressed. The excitation in modes
 $n\sim P$ is difficult to calculate, but should be of order one.
 The string mass (\ref{smass})
is then
\begin{equation}
\langle M^2_s\rangle =
8 \sum_{n=1}^{\infty} \tilde n \langle N_n\rangle \sim \sum_{n=1}^P
\tilde n \sim P^2
\end{equation}
where we have used the fact that for most of the terms, the excitation
is of order one.  (The first few terms contribute $P\ln^2 P$ which is
a subleading contribution.) Thus a string becomes very massive by the
time it crosses the horizon. Notice that this is the invariant rest
mass of the string, and not the kinetic energy seen by some observer.

We now check that the mass of the string remains much less than the
mass of the black hole.  Recall from (\ref{nsmass}) that the mass of
the black hole is
\begin{equation}
M \sim {r_0 RV\over g^2}(\cosh{2\gamma_1} + \cosh{2\gamma_p}) ,
\end{equation}
Under the reasonable assumption that $RV\ge 1$ and $g<1$, (\ref{P})
implies that $M> r_0^2 P$. In order for the
static curvature to be small near the horizon we need $r_0 \gg 1$, so
indeed $M \gg M_s$.

The typical size of the string can be estimated as follows
\cite{mit:stat}.  The mean squared radius of the string is roughly
\begin{equation}
\langle r^2 \rangle \sim \langle \int d\sigma :(x^i(0,\sigma) -
x^i_0)^2:\rangle 
\sim \sum {\langle N_n\rangle\over \tilde n}.
\end{equation}
Using the above estimate for the excitation yields
$\langle r^2 \rangle  \sim O(1)$, so despite the large mass,
the string remains small. 

We can also consider the process of Hawking radiation of
strings. Strings are much more likely to be radiated in their ground
state than in a highly excited state. But this refers to the state of
the strings near the horizon.  By the time they reach infinity, they
can be excited. If we model this process by starting the strings in
their ground state for $\tau < 1/P$, the Bogoliubov coefficients are
the same as for strings falling in.  Thus the excitation of the
$n^{th}$ mode in the asymptotic region will again be $\langle
N_n\rangle \sim \ln^2(n/P)$. The mass of the string in the asymptotic
region is given by the usual flat space formula, which again implies
$\langle M^2_s\rangle \sim P^2$.

\sect{Back-reaction of infalling matter}
\label{backr}

So far, we have discussed the behaviour of infalling matter in the
test particle (string) approximation. We now ask when this approximation 
breaks down. For the sake of simplicity, we will just consider infalling
dust in this section. Consider the contribution of the dust to
the stress tensor. The total stress tensor for the 
spacetime  is
\begin{equation} \label{stress}
T_{\mu\nu} = \bar T_{\mu\nu} + \rho u_\mu u_\nu, 
\end{equation}
where $\bar T_{\mu\nu}$ is the stress tensor for the background
spacetime, $\rho$ is the comoving density of the dust, and $u_\mu$ is
the tangent vector to the flow lines. For the test particle
approximation to be valid, the dust should follow the geodesics of the
background spacetime, and the backreaction of the dust on the
geometry, e.g., the change in the curvature invariants, should be
negligible. These conditions are always valid for sufficiently small
$\rho$. Naively, one might expect that $\rho^2 \ll \bar T_{\mu\nu}
\bar T^{\mu\nu}$ would be sufficient. However if we contract the total
stress tensor with itself, we obtain
\begin{equation} \label{cont}
T_{\mu\nu} T^{\mu\nu} =\bar T_{\mu\nu} \bar T^{\mu\nu} + 2\rho u^\mu u^\nu
\bar T_{\mu\nu} + \rho^2.
\end{equation}
We have seen that in the region of large tidal forces the geodesic
components of the Einstein tensor become much larger than the
curvature invariants.  This implies that the middle term on the right
in (\ref{cont}), $ 2\rho \bar T_{0'0'}$, can become larger than the
first, even if $\rho^2 \ll \bar T_{\mu\nu} \bar T^{\mu\nu}$. That is,
the presence of a small amount of infalling matter can give rise to
large curvature invariants in the region of large tidal forces, where
previously all the curvature invariants were small. Physically, this
is a result of a high energy collision between the infalling matter
and the background dilaton wave.

This raises the possibility that the behavior of infalling matter in
an exact solution will be qualitatively different from the test
particle approximation.  This would not contradict the known stability
of these black holes \cite{howi:stable} since this refers only to
linearized perturbations.  Fortunately, one can check this in a
special case involving the collapse of a thin shell.  In general
relativity, the motion of a thin shell of dust can be determined by
matching appropriate metrics on the inside and outside across the
shell, using the intrinsic matching conditions developed in
\cite{is:match,kuc:match}. In \cite{boul:coll}, Boulware used this
technique to describe the collapse of a spherically symmetric charged
dust shell to form a Reissner-Nordstr\"om black hole. The general
technique was extended to theories with gravity coupled to a scalar
field in \cite{bar:scalar,osh:scal}. In the case with a scalar field,
there is an additional requirement; the scalar field must be
continuous across the shell. This implies that if one tries to match
two static solutions across the shell, the position of the shell will
be forced to be static unless the scalar field has the same form in
the two solutions. Thus, we cannot study the collapse to form a
dilaton black hole this way, as the dilaton is constant in flat space,
and non-constant in the static black hole solution. Physically, a
collapsing spherical charged shell will radiate dilaton waves.  The
static black hole should be viewed as representing the geometry at
late times, long after the collapse has taken place.

However, if we wish to consider a spherical shell falling into an
existing black hole, we can use static solutions so long as the shell
carries no dilaton charge. This will imply that the dilaton has the
same form inside and outside the shell. We will consider the dilaton
black hole metric (\ref{gensol}), as this is the simplest case. The
metrics on the inside and the outside of the shell each have the form
(\ref{gensol}), with in general different values of $r_+, r_-$. (We
assume that the interior metric has large tidal forces outside the
horizon.) The position of the shell in the two metrics will in general
also be given by different functions $r(\tau)$, where $\tau$ is the
shell's proper time. However, $R(r)$ measures the proper area of the
shell, so it must have the same value at the shell in the two
metrics. The dilaton (\ref{dbhdil}) must also have the same value at
the shell in the two metrics. These two relations imply that if $a
\neq 0$, $r_-$ and $r(\tau)$ must both be the same in the two
metrics. The only difference is then in the value of $r_+$. We will
use $r_+$ to denote the value in the interior metric, and $r_+'$ to
denote the value in the exterior metric. The shell has a proper mass
$\cal M$ and charge $q$. The charge will be equal to the change in the
black hole's charge,
\begin{equation} \label{deltaq} 
q = \left({r_- \over 1+a^2}\right)^{1/2} (r_+'^{1/2} - r_+^{1/2}). 
\end{equation}
The black hole's mass changes by
\begin{equation} \label{deltam}
m = \frac{1}{2} (r_+' - r_+),
\end{equation}
which can be thought of as the total energy carried by the shell. This
total energy should be positive, so $r_+' > r_+$. Note that 
\begin{equation} \label{mqrel}
m  = {(r_+'^{1/2} + r_+^{1/2}) \over r_-^{1/2}} {\sqrt{1+a^2} \over 2}
\ q > \sqrt{1+a^2}\ q > q/\sqrt{1+a^2} = m_{BPS},
\end{equation}
so the total energy carried by the shell is greater than the BPS
bound. This is not necessarily true of its proper mass $\cal M$. Note
also that the requirement that the shell carries no dilaton charge
implies that its charge has the same sign as the black hole's.

The tangent vector to the shell is 
\begin{equation} \label{tshell}
u^\mu = (u^0, \dot{r}, 0,0),
\end{equation}
and the requirement that $u^\mu u_\mu = -1$ implies $u^0 = (F(r) +
\dot{r}^2)^{1/2} /F(r)$. The function $F(r)$ is discontinuous across
the shell, so the tangent will be different as seen from the inside
and the outside. We will use $F_i$ to denote the function that appears
in the interior metric, and $F_o$ to denote the function in the
exterior metric. The normal is orthogonal to $u^\mu$, and is hence
given by
\begin{equation} \label{norm} 
n_\mu =(-\dot{r}, u^0,0,0).
\end{equation}
The sign choices in the tangent vector and the normal have been made
so as to give a future-directed tangent vector and an outward-pointing
normal for a shell outside the black hole's event horizon. If we
define the notation $[K]_\pm \equiv K(r+\delta) - K(r-\delta)$ for the
discontinuity in some quantity $K$ across the shell, then the
intrinsic matching conditions of \cite{is:match,kuc:match} tell us
that the discontinuity in the extrinsic curvature is related to the
surface stress-energy of the shell, and hence \cite{boul:coll}
\begin{equation} \label{disc}
[u^\mu n_{\mu;\nu} u^\nu + n^\mu_{\phantom{\mu};\mu} - n^\mu n^\nu
n_{\mu ;\nu}]_\pm  = -2{\cal M}/R^2.
\end{equation}
As in \cite{boul:coll}, we can use the identity 
\begin{equation} \label{id}
\delta^\lambda_\sigma = \delta^\lambda_2 \delta^2_\sigma +
\delta^\lambda_3 \delta^3_\sigma + n^\lambda n_\sigma - u^\lambda
u_\sigma,
\end{equation}
and the fact that 
\begin{equation} \label{chris}
\Gamma^2_{\alpha 2} = \Gamma^3_{\alpha 3} = \delta^1_\alpha R'/R
\end{equation}
to simplify (\ref{disc}). The resulting equation is 
\begin{equation} \label{m1}
2{R' \over R} [n^1]_\pm = -2{\cal M}/R^2.
\end{equation}
Rearranging and inserting the values of the normals from (\ref{norm}),
we obtain
\begin{equation} \label{m2}
{\cal M}= R R'[ (F_i+\dot{r}^2)^{1/2} - (F_o + \dot{r}^2)^{1/2}]. 
\end{equation}
This equation determines the motion of the shells. Rearranging and
squaring twice to eliminate the radicals gives
\begin{equation} \label{m3}
\dot{r}^2 = \left[ {R R' \over 2 {\cal M}}(F_i - F_o) + {{\cal M}
\over 2 R R'} \right]^2 - F_i = \left[{m \over {\cal M}} \left( 1 -
{1 \over (1+a^2)}{r_- \over r} \right) + {{\cal M} \over 2R
R'}\right]^2 - F_i. 
\end{equation}

In the limit in which the mass of the shell goes to zero, (\ref{m3})
will reduce to the equation for charged test particles in the interior
metric. In this limit, $m \over {\cal M}$ will become the energy per
unit mass $E$ of the test particle, and the second term in the
brackets vanishes. This second term represents the contribution from
the self-gravity of the shell. Note that these test particle
trajectories are not geodesics, but describe the motion of charged
test particles with charge related to the energy by (\ref{mqrel}).
One can easily verify that the tidal forces experienced by these
charged test particles also becomes large in the region near the
black hole's event horizon.

The main point of this analysis is that the exact equation for the
motion of the shell (\ref{m3}) only differs from the corresponding
test particle equation by the self-gravity term, and hence the motion
of the shell is well approximated by test particles so long as this
term is negligible.  At large distances, this term can only be
important if $m = \cal M$ and $\cal M$ is of order $r_+$. This implies
that the shell starts at rest and that its mass is comparable to the
black hole's. However, we are more interested in whether this term can
become important at smaller radii.  Near the black hole, that is, at
$r = r_+ (1+ \epsilon \rho)$ ($\rho$ order one), $R R' \sim r_+
\epsilon^{(a^2-1)/(1+a^2)}$ where $\epsilon \equiv (1-r_-/r_+)$.  At
moderate distances, that is, $r = r_+ z$ ($z$ order one), $R R' \sim
r_+$. In both these regimes, the other term in the bracket is of order
one, so the ${\cal M}/2 RR'$ term will only be important if $\cal M$
is of the order of the smaller of $r_+$ and $r_+
\epsilon^{(a^2-1)/(1+a^2)}$.

We now wish to compare this condition on the mass of the shell
with the condition from the change in the curvature invariant
(\ref{cont}).  In the region of large curvatures, the second term in
(\ref{cont}) is
\begin{equation} \label{rhoG}
2 \rho \bar T_{0'0'} \sim {{\cal M} \over l R^2} {1 \over r_+^2 \epsilon^2}
\sim {{\cal M} \over l r_+^4 \epsilon^{2(1+2a^2)/(1+a^2)}},
\end{equation}
where $l$ is the comoving thickness of the shell, which for simplicity
we will now assume to be of order one\footnote{Since the radial tidal
forces remain small, $l$ is approximately constant along the
trajectory.}.  Now this term will be of order one (in Planck units) if
${\cal M}$ is of order $r_+^4 \epsilon^{2(1+2a^2)/(1+a^2)}$. We can
choose the black hole's parameters so that the shell's self-gravity is
still negligible at this value. Thus, the test particle approximation
used in computing (\ref{rhoG}) is valid. So the curvature invariant
can become as large as the Planck scale (which is certainly much
bigger than its background value) without a noticeable departure from
the test particle trajectories.

The curvature invariant is only large in the shell, where $\rho\ne 0$.
Since we have used a thin shell approximation where the stress energy
of the shell is essentially a $\delta$-function, one might wonder if
this is responsible for the large invariant. This is certainly {\em
not} the case.  The invariant we compute is small far from the black
hole, and was estimated assuming a finite thickness $l$ for the
shell. Although we chose $l$ to be of order the Planck length, this
was not essential.  The point is simply that the curvature invariant
(\ref{rhoG}) and the self-gravity term in (\ref{m3}) depend on the
black hole parameters in different ways, so for any $l$ we can choose
the parameters so as to make the curvature invariant as large as we
want, while still keeping the self-gravity term negligibly small.

Not only does the large curvature invariant have little effect on the
motion of the infalling matter, it also has little effect on the
subsequent evolution of the spacetime. After the shell passes, the
geometry is again a static charged black hole with mass and charge
increased by the shell.  The original black hole was near extremal
with the mass above extremality of order $\Delta M \sim r_+
\epsilon$. Remarkably, one can fine-tune the black hole parameters so
that the curvature invariant (\ref{rhoG}) is large while keeping
${\cal M} / \Delta M$ small. In other words, one can achieve Planck
scale curvature invariants with only a small change in the background
metric.

\sect{Discussion}
\label{concl}

 We have seen that the large tidal forces found for certain
near-extreme black hole solutions in \cite{hor:naked} have a simple
physical interpretation. They result from a large dilaton wave
traveling just outside the event horizon\footnote{As we have seen,
this wave is an approximate description near the horizon. Since the
solution is static, it does not carry away any energy.}.  This wave
might be viewed as the remnant of the collapse to form the black
hole. We saw in the last section that the collapse of a spherical thin
charged shell in theories with a dilaton results in the emission of
scalar waves. The static black hole solution should be viewed as
representing the geometry at late times, long after the collapse has
taken place. For near extremal black holes, even after everything has
settled down, dilaton waves remain hovering just outside the
horizon.  The fact that the tidal forces are directly related to the
proper time remaining to the singularity seems to imply that the wave
remains significant only for near-extreme black holes.

When a test body falls into a naked black hole, it is crushed by the
large tidal forces. The geometry near the horizon seen by an infalling
observer can be approximately described by a dilaton plane wave
metric. We applied this simplifying approximation to show that when a
test string falls in, it becomes highly excited. However, the average
size of the string does not grow significantly relative to a fixed
length scale as it falls into the black hole.

When one includes the stress energy of the infalling matter, we have
seen that the curvature invariants become large near the
horizon. However, this does not lead to any significant change in the
classical solution. This surprising fact can be understood as
follows. Consider the total stress tensor (\ref{stress}) in the
geodesic frame. Only the $T_{0'0'}$ component will be different from
the background. The background component $\bar T_{0'0'}$ is large and
of the same order as the tidal forces, while the correction is just
$\rho$, and hence much smaller than the background.  Since there is a
frame in which the components of the stress tensor receive only small
corrections, the change in the metric must also remain small. The fact
that the curvature invariants can still become large is perhaps best
illustrated by a simple vector analogy.  Consider a large, nearly
lightlike vector; that is, one whose time and space components in some
orthonormal basis are individually large, say of order $X \gg 1$, but
whose norm is very small. If we add to it a small vector which only
has a time component, of order $1/X$, the resulting vector is close to
the initial one, but its norm is now of order one.

Since infalling matter can produce Planck scale curvature invariants
outside the horizon, quantum effects should become important in this
regime. It is even possible that quantum effects are important for
naked black holes before infalling matter is added. This cannot take
the form of perturbative local corrections since all curvature
invariants are small. But nonlocal (or nonperturbative) effects might
be important. This issue certainly deserves further
investigation. This is especially true in light of the recent ideas in
\cite{jac:decay}, in which it was argued that Hawking radiation from a
near extremal Reissner-Nordstr\"om black hole may cause the horizon to
become unstable. The dilaton wave near the horizon of naked black
holes might be viewed as a classical analog of the Hawking radiation
in the Reissner-Nordstr\"om case.

\bigskip
\bigskip
\centerline{\bf Acknowledgments}
\medskip
It is a pleasure to thank S. Giddings, T. Jacobson and E. Weinberg for
discussions. This work was supported in part by NSF grant
PHY95-07065. SFR also acknowledges support from the Natural Sciences
and Engineering Research Council of Canada.

\def\href#1#2{#2}

\begingroup\raggedright\endgroup

\end{document}